# Magnetism and Verwey transition in magnetite nanoparticles in thin polymer film

V.N. Nikiforov[1,*], Yu.A. Koksharov[1], S.N. Polyakov[1], A.P. Malakho[2],
A.V. Volkov[3], M.A. Moskvina[3], G.B. Khomutov[1], V.Yu.Irkhin[4]

[1]Faculty of Physics, M.V.Lomonosov Moscow State University, 117234, Moscow, Russia
[2]Department of Material Sciences, M.V.Lomonosov Moscow State University, Moscow, Russia
[3]Faculty of Chemistry, M.V.Lomonosov Moscow State University, Moscow, Russia
[4]Institute of Physics Metals, Ekaterinburg, Russia
*E-mail: nvn@lt.phys.msu.ru

**Abstract.**
Magnetic and structural properties of magnetite nanoparticles stabilized in polyvinyl-alcohol thin films are investigated by using X-ray diffraction (XRD), transmission electron microscopy (TEM), electron paramagnetic resonance (EPR) and static magnetometry techniques. The nanoparticles have well-defined crystallinity, and are superparamagnetic at room temperature. Their size distribution is characterized by the distinct log-normal law (with average diameters near 5-7 nm) and slight maximum near 70-80 nm. The EPR spectra and static magnetization data demonstrated pronounced anomalies in the interval between 130 K (corresponding to Verwey transition) and 200 K. The experimental data obtained can be understood on the basis of the half-metallic electronic structure, complex temperature behavior of the magnetic anisotropy, along with effects of "weak magnetic-electron" sublattice of the magnetite.

*Key words: nanoparticles, magnetite, EPR, magnetization*

## 1. Introduction

Magnetite $Fe_3O_4$ is one of most wide-spread natural iron compounds and the most ancient known magnetic substance. However, we still have no complete explanation of its magnetic, electronic and even structural properties.

Nano-sized magnetite is very important material which is used for various applications including ferrofluids synthesis, drug delivery *etc*. Magnetite nanoparticles also play important role in living organisms, e.g. in magnetotactic bacteria, where highly organized chains of magnetite nanocrystals provide a possibility for their navigation in the Earth's magnetic field [1]. Bulk magnetite is a ferrimagnetic spinel with a high Curie temperature of 858 K At $T_V \approx 125$ K, the bulk $Fe_3O_4$ shows electrical, magnetic and other anomalies, that indicates a phase transition known as the Verwey transition. Also $Fe_3O_4$ has an isotropic point at 130 K where the first magnetocrystalline anisotropy constant $K_1$ passes through zero. Magnetic properties of bulk magnetite are still far from full understanding [2]. For example, various magnetic anomalies (magnetic after-effects [3], non-monotonous behavior of $K_1$ [4], etc.) were found for bulk magnetite in the temperature interval 200-300 K. There is no consensus about the significance of magnetite magnetic and electron subsystems in the Verwey transition origin.

Nanoparticles usually demonstrated changes in magnetic and other properties in comparison with the bulk counterparts, which can be due to the finite-size and surface effects, as well as to structure and morphology modifications [5].

Earlier, superparamagnetism was detected by using Mössbauer spectroscopy and magnetization measurements in small (about 10 nm in size) magnetite and maghemite particles embedded in thin (100-200 μm) polyvinyl alcohol polymer films [6, 7]. In the present work, we report in detail on structural and magnetic studies of $Fe_3O_4$ nanoparticles. These nanoparticles are substantially smaller in comparison with those in conventional ferrofluids and are characterized by a distinct log-normal size distribution and well-defined crystallinity. We pay especial attention to properties near the Verwey transition.

**2. Experimental methods**

The nanoparticle synthesis method was based on *in situ* reactions taking place in the bulk of the polymer matrices which undergo swelling in water. Details of the preparation technique were described in the papers [6,8]. We studied a set of samples with different (from 0.6% to 3%) magnetite volume concentration. The weight concentration in all the samples was equal to approximately 5%. The particle-size distribution was studied by the TEM spectroscopy using JEOL JEM-100B microscope. X-ray diffraction measurements were performed with the Rigaku Rotaflex D/max-RC diffractometer (X-ray wavelength $\lambda = 0.15405$ nm (Cu $K_\alpha$), the power of X-ray source being equal to 12 kW. The EPR spectra were recorded with the computerized spectrometer Varian E-4 (X-band) and the flow-nitrogen temperature controller E-257 at temperatures from 100 K to 300 K. The static magnetization was studied by the vibrating sample magnetometer PARC-M-153 as well as by a home-made SQUID at temperatures from 4 K up to 300 K.

**3. Results**

In general, the structural and magnetic properties of the samples studied are not strongly dependent on the magnetite volume content in the polymer matrix. The TEM pictures for all the samples showed that the nanoparticles are distinctly divided in two groups: rather small (below 10 nm) and relatively large (about 70-80 nm). The representative TEM photograph for the sample with 0.6% vol. of $Fe_3O_4$ is shown in Fig.1a. The log-normal size distribution is typical for nanoparticle systems. The XRD spectra of $Fe_3O_4$ nanoparticles showed well-resolved diffraction peaks corresponding to the structure of the bulk magnetite (Fig.1b). The average size of crystallites, according to the XRD data, is equal to 5-7 nm. That is, in our samples the particles with sizes less than 10 nm are dominated. This conclusion is confirmed also by the absence of the magnetic remanence for all the samples at room temperature.

Fig.2a shows the temperature dependences of the static magnetic moment $M$ measured by SQUID during zero-field-cooling (ZFC) and field-cooling (FC) procedures for the sample with 1.2% vol. of $Fe_3O_4$. The rather broad maximum of the ZFC curve is placed near 175 K, and the ZFC and FC divergence takes place at $\approx 300$ K. Fig.2b presents the ZFC curve measuring by the VSM for the same sample. It follows from the comparison of Figs.2a and 2b that the increase of the measuring field from 45 Oe (SQUID) up to 1 kOe (VSM) results in significant narrowing of the maximum of the ZFC curve and the shift of this maximum to markedly lower temperatures. The temperature dependence of the $M_{ZFC}$, presented in Fig.2b, demonstrates rapid growing of the magnetic moment near 130 K. This phenomenon is absent in low-field magnetization measurements (Fig.2a).

The EPR characteristics also demonstrate some anomalies in their temperature behavior. For example, Fig.3 shows how the EPR peak-to-peak linewidth $\Delta H_{pp}$ monotonously increases in general with the temperature decrease, but the slope of the curve $\Delta H_{pp}(T)$ changes near 140-150 K. The effective resonance magnetic field $H_{res}$ shows the same tendency to change the slope of the temperature dependence. Below 140 K lowering of $H_{res}$ with decrease of temperature becomes more pronounced (Fig. 4a and 4b).

The non-monotonic temperature behavior was found also for the EPR intensity $I_{EPR}$, which was obtained by the double integration of the experimental spectra. Fig. 5 presents the temperature dependences of $I_{EPR}$ for the samples with 0.6 and 3% vol. of $Fe_3O_4$ concentration. In both cases the normalized EPR intensity has a broad maximum at 170-200 K (Fig.5). When the temperature decreases from 300 K down to 100 K, the swing of the $I_{EPR}$ relative variation amounts to $\approx 10\%$.

**4. Discussion and conclusions**

Usually the phase transition in magnetite at $T_V \approx 125$ K, accompanied by a structural distortion, is treated as a charge ordering of $Fe^{2+}$ and $Fe^{3+}$ states. However, the nature of the Verwey transition and low-temperature phase is still a subject of numerous investigations. Electronic structure calculation [9] demonstrate that magnetite in the cubic spinel structure provides an example of half-metallic ferromagnetic state with majority-spin gap (see discussion in the review [10]). This means a magnetically saturated state of itinerant 3d electrons propagating over B-sites in a narrow strongly-correlated band, the magnetic moment being close to $4\mu_B$ per formula unit.

This picture was questioned by the x-ray magnetic circular dichroism (XMCD) data [11] interpreted as an evidence of large orbital contributions. However, later XMCD experiments [12] confirmed the purely spin saturated magnetic state. Direct measurements of spin polarization by spin-polarized photoemission spectroscopy [13] yield the value about -40% only (instead of –100% predicted by simple band picture). Therefore the surface effects can be expected to influence strongly magnetic and electronic properties (in particular, the Verwey transition), especially for nanoparticles.

It is known for the bulk magnetite that the development of the Verwey transition is sensitive to structural peculiarities of a sample, such as oxygen stoichiometry, vacancy and cation distribuition, etc. [4]. It should be all the more true for magnetite nanoparticles, since, as a rule, nano-size particles have usually even more complex structure than their bulk counterparts [5, 14]. However, our XRD data show rather good crystalline structure for the nanoparticles studied, close to that of the bulk magnetite. Besides, the non-stoichiometry shifts the Verwey transition temperature to the lower values [4]. Hence, the static magnetization and EPR anomalies observed in our work should be rather addressed to intrinsic magnetite properties.

A possible reason of magnetic anomalies in magnetite nanoparticles could be the inter-particle interactions. But the $Fe_3O_4$ concentration in the polymer matrix is rather low, and the nanoparticles are well separated from each other (Fig.1a). Besides, the temperature dependences observed are hardly typical of collective effects like magnetic ordering.

The ZFC and FC curves presented in Figs.2a and 2b look typical for nanoparticle systems [14]. The temperature $T_p$ of the peak in $M_{ZFC}(T)$ curve, recorded at low measuring magnetic fields, is usually roughly identified with the average blocking temperature $<T_B>$, though the $T_p$ value can be significantly influenced by the particle size and shape distributions, the cooling rate during magnetization recording, inter-particle interactions and many other factors [14-16].

The ZFC magnetization curve, shown in Fig.2a, is characterized by rather broad and flat plateau. Such unusual shape can be due to the superposition of curves with different peak points and Curie temperatures, corresponding to groups of particles with the average sizes ~ 7 nm and ~80 nm. The first group is described by distinct log-normal law. The number of latter type (heavy) nanoparticles is small, so that they can be hardly observed in the size distribution (only slight maximum is seen), but give an appreciable contribution to magnetization.

Unusual ZFC magnetization pattern can be connected with anomalous temperature behavior of the magnetic anisotropy (see below). The application of the higher magnetic field $H$ affects the shape of the $M_{ZFC}(T)$ curve and shifts its peak to low temperatures (Fig.2b). Assuming that the $T_p$ value obeys the power law, $T_p \sim (1 - H/H_a)^2$, $H_a$ being the anisotropy field [15], we have estimated the value $H_a$ as $2 \cdot 10^3$ Oe. Since $H_a = 2K/M_S$, where $M_S$ is the saturation magnetization, the anisotropy constant $K$ can be evaluated. Assuming the bulk zero-temperature value $M_{S,b}$ = 92 emu/g, we get $K \approx 10^5$ erg/cm$^3$.

The magnetization data presented in Fig.2b show the magnetic moment which is much smaller than $M_{S,b}$. It means that the measuring magnetic field of 1 kOe is far from the saturation value. In magnetite nanoparticles with sizes of $\approx$ 120 nm the relative saturation of the magnetization (at the rate of about 60% of $M_{S,b}$) was observed only at 20 kOe [16].

The value of magnetic anisotropy, obtained for the magnetite nanoparticles, is close to the low-temperature bulk value [17]. This seems to be in contradiction with the usual tendency of significant increasing of the magnetic anisotropy constant in nanoparticles in comparison with the bulk counterparts [5]. On the other hand, it is well known that magnetic anisotropy in the bulk magnetite depends strongly on the temperature in the interval 130 K-200 K [4,18]. For example, the first magnetocrystalline anisotropy constant $K_1$ passes through zero at 130 K [18]. The lowering of the EPR resonance field below 130-150 K, found for the magnetite nanoparticles in our work (Fig.4), can be also due to the decreasing of the magnetic anisotropy below 130-150 K. Indeed, this decrease of the magnetic anisotropy results in increasing of magnetic susceptibility, which was really observed in bulk magnetite above the Verwey transition [19] and probably can account for the anomaly of the ZFC magnetization at 130 K in our samples (Fig.2b).

The anomalies of magnetic properties of the bulk magnetite above the Verwey temperature are not new phenomena. Magnetic after-effects were observed in the temperature interval 200-250 K [3]. It was stressed by Belov [19] that the spontaneous magnetization $\sigma_S$ above $T_V$ has a broad maximum near 200 K. At the same temperature interval the magnetic paraprocess demonstrates the increasing of the susceptibility. According to the Belov model, the "magnetic-electron" subsystem ("weak magnetic-electron sublattice") in magnetite shows unusual magnetic behavior well above the Verwey transition, starting from nearly 300 K and finishing below 100 K. It is interesting that the shape of $I_{EPR}(T)$ in our samples (Fig.5) is similar to that of $\sigma_S(T)$ in bulk magnetite above $T_V$ [17]. The theoretical estimations of relative change in the "magneto-electron" subsystem contribution to $\sigma_S$ above $T_V$ give the value about 20% [19]. This is close to the change in the EPR intensity in the temperature interval 130-300 K (Fig.5).

To summarize, we found pronounced anomalies in the static magnetization and EPR behavior of magnetite nanoparticles at temperatures near and well above the Verwey transition. We suppose that these anomalies are concerned either with the complex behavior of the magnetic anisotropy in the magnetite or with peculiar properties of the "magnetic-electron" subsystem.

This work was supported by Russian Foundation for Basic Researches (Grant 02-03-33158) and INTAS (grants 99-1086, 01-483), by the Programs of fundamental research of RAS Physical Division "Strongly correlated electrons in solids and structures", project No. 12-T-2-1001 (Ural Branch) and of RAS Presidium "Quantum mesoscopic and disordered structures", project No. 12-P-2-1041.

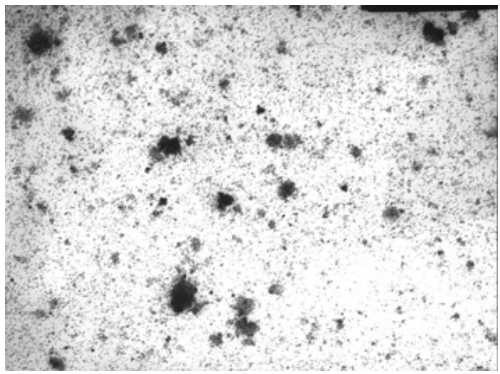
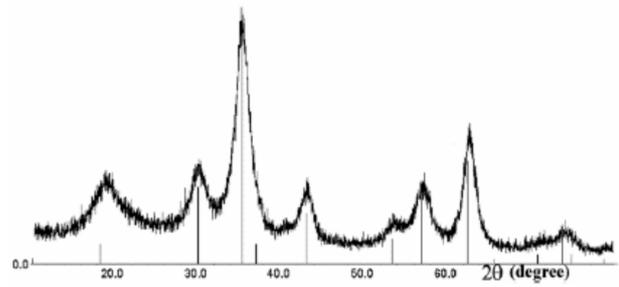

(a)                                       (b)

Fig.1 (a) TEM pictures for the sample with 0.6% vol. of $Fe_3O_4$. The picture width corresponds to 1000 nm. (b) X-ray diffraction spectra for the sample with 1.2% vol.

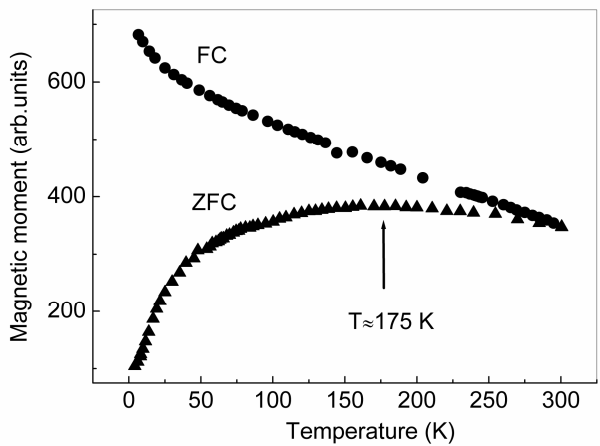
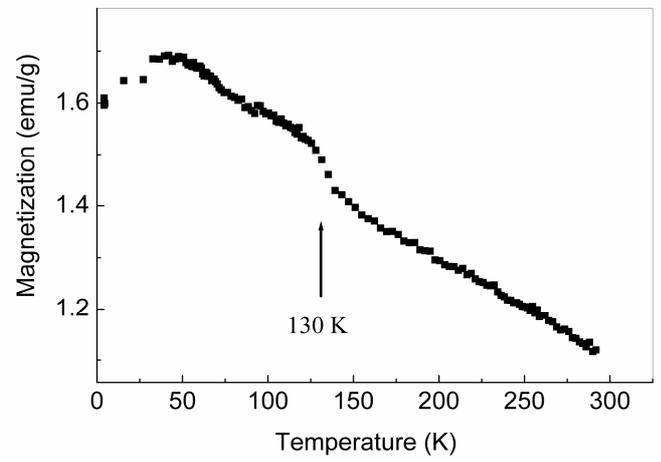

(a)                                       (b)

Fig.2 The temperature dependencies of the magnetic moment for the sample with 1.2% vol. of $Fe_3O_4$. (a) ZFC and FC curves measured by the SQUID. The measuring (during ZFC and FC processes) and the cooling field (during FC process) is equal to 45 Oe. (b) The ZFC curve measured by the VSM. The measuring field is 1 kOe.

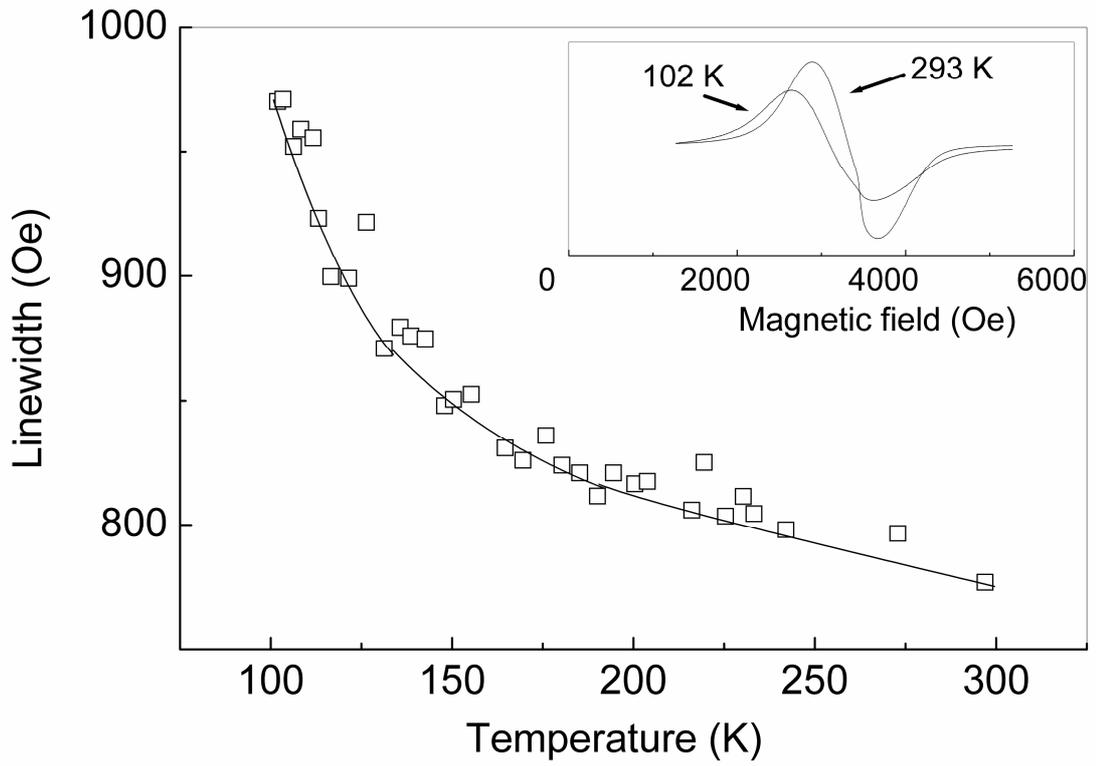

Fig.3. The temperature dependence of the EPR linewidth for the sample with 0.6% vol. of $Fe_3O_4$. The external magnetic field is parallel to the film plane. Solid line is the guide for eye. The insert shows the typical EPR spectra at different temperatures.

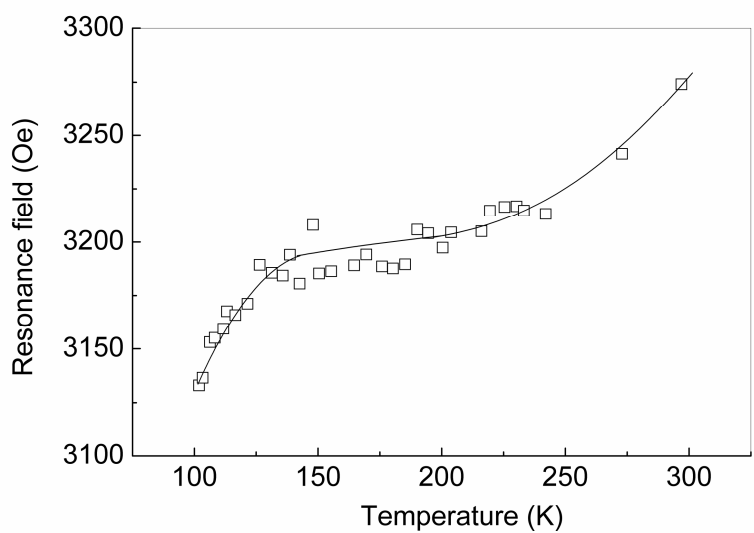 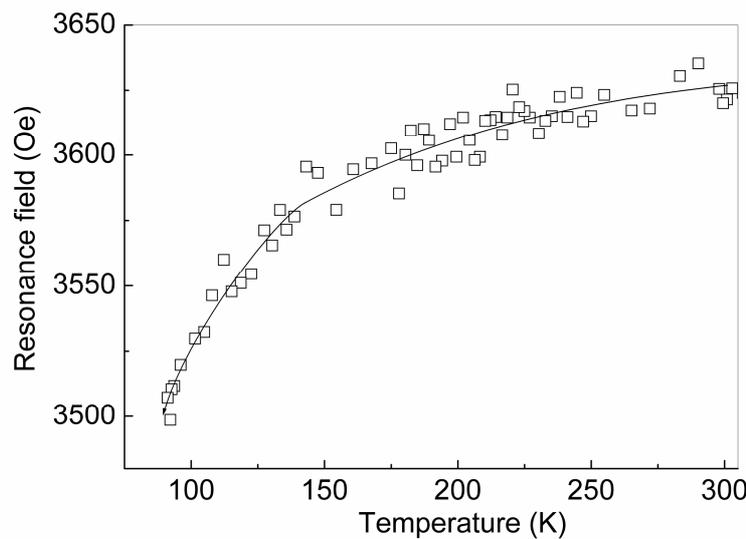

(a)                                                      (b)

Fig.4. The temperature dependence of the EPR resonance field. Solid lines are the guides for eye. (a) The sample with 0.6% vol. of $Fe_3O_4$. External magnetic field is parallel to the film plane. (b) The sample with 1.2% vol. of $Fe_3O_4$. External magnetic field is perpendicular to the film plane.

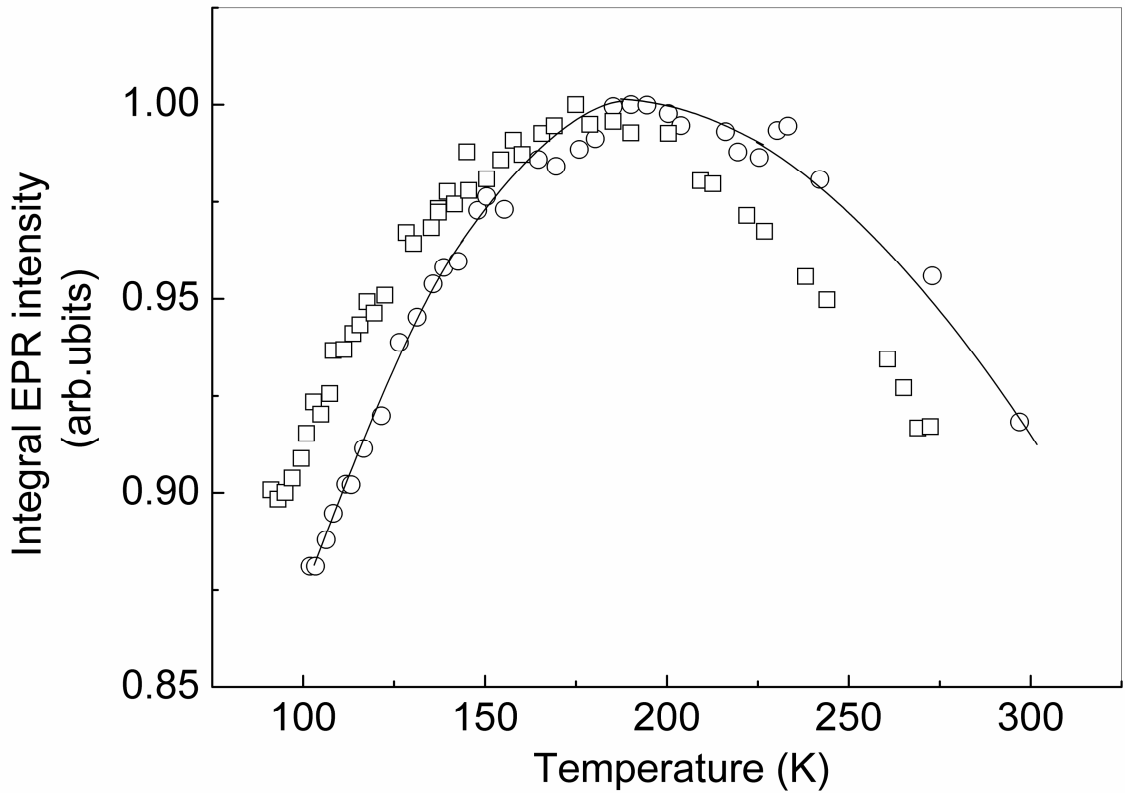

Fig.5. The temperature dependence of the integral EPR intensity. Circles correspond to the sample with 0.6% vol. of $Fe_3O_4$. The external magnetic field is parallel to the film plane. Squares stand for the sample with 3% vol. of $Fe_3O_4$. The external magnetic field is perpendicular to the film plane. Solid line is the guide for eye.